# Secure Internet Exams Despite Coercion

Mohammadamin Rakeei[1], Rosario Giustolisi[2], and Gabriele Lenzini[1]

[1] SnT, University of Luxembourg, Esch-sur-Alzette, Luxembourg
[2] Department of Computer Science, IT University of Copenhagen, Denmark
{amin.rakeei,gabriele.lenzini}@uni.lu, rosg@itu.dk

**Abstract.** We study coercion-resistance for online exams. We propose two properties, Anonymous Submission and Single-Blindness which, if hold, preserve the anonymity of the links between tests, test takers, and examiners even when the parties coerce one another into revealing secrets. The properties are relevant: not even Remark!, a secure exam protocol that satisfied anonymous marking and anonymous examiners results to be coercion resistant. Then, we propose a coercion-resistance protocol which satisfies, in addition to known anonymity properties, the two novel properties we have introduced. We prove our claims formally in ProVerif. The paper has also another contribution: it describes an attack (and a fix) to an exponentiation mixnet that Remark! uses to ensure unlinkability. We use the secure version of the mixnet in our new protocol.

## 1 Introduction

One of the most tangible consequences of the Corona virus pandemic in education has been that academic institutions moved exams to the Internet. The migration to an online format did aggravate the problem of fraud, which is now a concern for all institutions worldwide (Watson, 2010).

Cheating at exams, *i.e.,* taking advantage of the process for one's own benefit, is a practice old as the establishment of exams, and it is unsurprising that candidates try to cheat; However, the use of Information and Communication Technologies (ICT) (called *electronic exams*, or e-exams) makes it easier and more appealing for examiners and administrators to temper with the exam processes. Authorities have been found to change grade records to boost their university national ranking [3]; More recently, a famous legal firm has been fined 100 million USD because hundreds of auditors at the firm cheated at ethics tests required to keep their professional license [4].

---

[3] Valerie Strauss, "Remember the Atlanta schools' cheating scandal? It is not over", 1 February 2022, Washington Post, accessed on 2022/06

[4] Tory Newmyer, "Ernst & Young hit with a 100 million fine over cheating on ethics tests", 28 June 2022, Washington Post, accessed 2022/07

The problem with fraud in electronic exams is that the implementation of traditional mitigation actions is harder and less scalable than in traditional pencil-and-paper exams. Frauds perpetrated by attacking the underlined communication infrastructure are also subtle and hard to detect. Migrating exams online may compromise the quality of education assessment unless we can provide exam protocols that are *secure by-design*.

Security in electronic exams is not a new topic. Properties such as anonymous marking, question secrecy, and mark integrity have been studied and formally expressed (Giustolisi, 2018); e-exam protocols that are secure under specific threat models, for example with curious authorities, have been proposed (Bella et al., 2017);. Researchers have discussed e-exam protocols that are *verifiable* (Dreier et al., 2014), and *privacy-preserving verifiable* (Giustolisi et al., 2017).

*Contribution* How this work contributes to this ensemble of works on the security of e-exams? First, it studies two new properties meant to capture an important class of security requirements, that of *coercion resistance*, never studied in the context of e-exams. The new properties are called *Anonymous Submission* and *Single-Blindness*. The former expresses that examiners cannot get to know the link between test takers and the tests they submitted, even if examiners force candidates to reveal their secrets; the latter says that candidates cannot learn the link between examiners and the tests they have marked, even if they compel examiners to reveal their secrets.

One can ask whether the two new properties are indeed novel. After all, they sound similar to *Anonymous Marking* (marks remain secret until notification) and *Anonymous Examiner* (test takers ignore the identities of examiners). To answer the question, we select a state-of-the-art secure exam protocol, Remark! (Giustolisi et al., 2014), which uses the exponentiation mixnet proposed by Haenni-Spycher (Haenni and Spycher, 2011) to preserve anonymity and unlinkability and that satisfies *Anonymous Marking* and *Anonymous Examiner* under the assumption that the roles are malicious and can collude. Under collusion, Remark! does not satisfy *Anonymous Submission* and *Single-Blindness*, giving the argument that the new properties express something unprecedented.

On this line, we propose a new *coercion-resistant exam protocol*, which we call Coercion-Resistant Electronic Exam (C-Rex). It satisfies *Anonymous Submission* and *Single-Blindness* under coercion, and we prove this statement formally using Proverif (Blanchet, 2001), a famous model checker. We also verify that C-Rex satisfies the other security properties already met by Remark!.

This work has another orthogonal, but not less important, contribution. We discover that by injecting specific messages into *exponentiation mixnet* used by Remark! to ensure unlinkability, an attacker can break unlinkability. An immediate consequence is that, in Remark!, various anonymity properties no longer hold, including *Anonymous Marking* and *Anonymous Examiner*. We propose a fix to the original exponentiation mixnet that, in Remark! restores *Anonymous Marking* and *Anonymous Examiner*. We believe that our fix also improves the security of Haenni-Spycher's mixnet in other domains than the online assessment, a statement that will be investigated in future work.

## 2 Related Work

Although a discussion about security in e-exams has recently gained attention, the topic is not new and has been researched before. For a fairly extensive list of requirements for e-exams, we refer the reader to (Giustolisi, 2018); for a brief account, we recall that security requirements for e-exams have been informally expressed in (Weippl, 2005; Furnell et al., 1998), while (Dreier et al., 2014) defines a formal framework in the applied $\pi$-calculus to specify and analyze authentication and privacy in e-exams; in addition, the authors define and show how to prove a set of verifiability properties for exams (Dreier et al., 2015). In a similar domain, that of computer-supported collaborative work, (Foley and Jacob, 1995) formalizes confidentiality, proposing exams as a case study.

In the state-of-the-art, we find not only works that describe properties for secure and private assessment, but also proposals for new security protocols for computer-assisted exams. (Castella-Roca et al., 2006) designs a protocol that meets authentication and privacy properties in the presence of a fully trusted exam manager. (Bella et al., 2014) proposes an e-exam, which considers a corrupted examiner, but assumes an honest-but-curious anonymiser. This assumption was later removed in (Bella et al., 2017).

(Huszti and Pethö, 2010) tackles on-line exams and discusses an Internet-based exam protocol with few trust requirements on principals. (Giustolisi et al., 2014) describes *Remark!*, another Internet-based exam protocol that ensures authentication and conditional anonymity requirements with minimal trust assumptions; this is the protocol we refer to here as use case.

The interest in e-exams is not limited to researchers. Many institutions for language proficiency tests, for example the Educational Testing Service (ETS) [5], or organizations for personnel selection, for instance Pearson[6] or the EU EPSO [7], offer computer-assisted and online testing.

Some related protocols have been proposed in the area of conference management systems. In fact, our proposed properties can be seen as double-blind review properties, typically requested in the submission of conference papers. (Kanav et al., 2014) introduced *CoCon*, a formallr verified implementation of a conference management system that guarantees confidentiality. Arapinis *et al.* introduced and formally analysed *ConfiChair*, a cryptographic protocol that addresses secrecy and privacy risks coming from a *malicious-but-cautious* cloud. Their work has been extended to support any cloud-based system that assumes honest managers, such as the public tender management and recruitment process (Arapinis et al., 2013).

All these previous works do not mention or consider coercion resistance, a class of properties that, instead, we study here for e-exams.

---

[5] https://www.ets.org/

[6] https://www.pearsonassessments.com/

[7] https://epso.europa.eu/

## 3   Background

Exam procedures differ one another in the details, but they all share the same organization and information flow at a certain level of abstraction. They are generally organized in four distinct phases: *Registration*, *Examination*, *Marking*, and *Notification*. Several roles are involved: *candidates*, which are the test takers; *question committee*, which prepares the exam questions; *examiners*, those who mark the tests; and *exam authorities*, a set of potentially distinct agents that help with the management *e.g.,* to distribute the test to students, assign examiners, and notify marks.

### 3.1   Security Properties of an e-Exam

An e-exam protocol is expected to guarantee specific security properties. The ones this work refers to are: (i) *Test Answer Authentication* (only tests submitted by eligible candidates are collected); (ii) *Examiner Authentication* (only authenticated examiners are allowed to marking tests); (iii) *Mark Privacy* (the marks given to an exam remain unknown to the other candidates); (iv) *Anonymous Marking* (examiners do not learn the identity of the candidate who submitted a test until after the test is marked); (v) *Anonymous Examiner* (candidates do not learn the identity of the examiners who marked their tests).

To this list, this paper adds *Anonymous Submission* and *Single-Blindness*, which are defined below.

### 3.2   Coercion resistant e-Exams

Among the properties that exist about e-exams, no one refers to coercion-resistance. In layman terms, this means that an exam protocol should preserve privacy between examinees and examiners even in case one of the parties tries to force to other into reveal secrets that could lead to de-anonymization.

The threat is not negligible considering that e-exams are not only used for university grading but also for general skill assessment, for instance in national educational assessments like the PISA[8], the PIAAC[9] exercise, and the various Tests for English as a Foreign Language required to get VISA and studentships, like the TOEFL family. Here, the benefits at stake are higher and so are the incentives to stress the procedure for one's own profit.

To address coercion resistance, we propose two additional security requirements for e-exams, *Anonymous Submission* and *Single-Blindness*. The former states that an examiner cannot learn which answer is submitted by a candidate. At first glance, *Anonymous Submission* sounds similar to *Anonymous Marking*, as both aim to achieve unlinkability between the identity of a candidate and the answer that she submitted. The difference is that the definition of *Anonymous Submission* is based on the indistinguishability of the answers submitted by two known candidates, while the definition of *Anonymous Marking* is based on the indistinguishability of the candidates' identities who submitted two known answers. As shall we see in Section 5.1, this is a key difference that allows us to model coercion in an e-exam by starting from the definition of *Anonymous Submission*. The property is relevant when an examiner tries to coerce candidates to reveal their private keys, or the answers they submitted.

---

[8]The Programme for International Student Assessment

[9]Programme for the International Assessment of Adult Competencies

The latter, that is *Single-Blindness*, ensures that no candidate can learn which mark has been assigned by an examiner. This property recalls *Anonymous Examiner*, as they both aim to achieve unlinkability between the identity of an examiner and the mark provided by this: however, in this case, the difference between the properties is neat. While the definition of *Anonymous Examiner* is based on the indistinguishability of the examiners' identities who provided a mark for two known tests, *Single-Blindness* is based on the indistinguishability of the marks provided by two known examiners. Here, the main difference is that *Single-Blindness* does not consider the tests at all. The property is relevant when a candidate attempts to compel an examiner to reveal a private keys to get access to the marks he has reported to the exam authority.

## 4 Use Case and a Relevant Attack on Mixnet

We are interested to verify whether our new properties are relevant. We chose as the use case a state-of-the-art exam protocol Remark! (Giustolisi et al., 2014). The reason for this choice is that Remark! is an internet-based exam that preserves all the properties listed in the previous section, including *Anonymous Marking* and æ, without assuming trusted third parties. All parties are malicious and can collude. It seems a good choice to start with to explore resilience against coercion.

### 4.1 Remark! in a nutshell

This online exam protocol is organized into typical phases, as we explained earlier. At *Registration* private/public keys and pseudo-identities are generated and distributed to candidates and examiners using an exponentiation mixnet; during *Examination*, questions are anonymously distributed (relying on a bulletin board) to candidates; the exam authority receives the answers submitted by the candidates; at *Marking* the æanonymously distributes the tests to examiners; at *Notification*, the exam authority records the marks and notifies the candidates their results. Dreier *et al.* in (Dreier et al., 2014) formally proved that Remark! satisfied *Test Answer Authentication*, *Mark Privacy*, *Anonymous Marking* and *Anonymous Examiner*.

Remark! relies on the Haenni-Spycher's exponentiation mixnet (Haenni and Spycher, 2011) to generate pseudo-identities for examiners and candidates.

### 4.2 Haenni-Spycher's Exponentiation Mixnet

(Haenni and Spycher, 2011) proposed a structure called *verifiable shuffling mixnet* or *exponentiation mixnet*, and designed an e-voting protocol based on it. The aim of the mixnet is to generate a list of pseudonyms from a list of public keys in a way that only the owner of a public key can find the corresponding pseudonym in the output list. This mixnet construction is used in Remark! to output a list of pseudo-public keys (pseudonyms) from an input list of public keys, while preserving unlinkability between input and output lists. To do so, exam authority sends the list of eligible candidates' and examiners' public keys. to the mixnet. Upon receiving the list of public keys the mixnet generates the output list as depicted in Figure 1. In this structure, $PK_i = g^{SK_i}$, $\overline{PK}_i = h_C^{SK_i}$, $\overline{r}_k = \prod_{i=1}^{k} r_i$ and $\overline{\pi}_k = \pi_k \circ \cdots \circ \pi_1$, where $SK_i$, $r_k$ and $\pi_k$ are, respectively, the private key of the party $i_{th}$, the secret element of the mixnet $k_{th}$ and the secret permutation of the mixnet $k_{th}$. For a more detailed description, see (Haenni and Spycher, 2011).

$$
\begin{array}{c|cccc|}
& mix_1 & mix_2 & & mix_m \\
\hline
C_1 \; PK_1 & PK_{\overline{\pi}_1(1)}^{\overline{r}_1} & PK_{\overline{\pi}_2(1)}^{\overline{r}_2} & \cdots & PK_{\overline{\pi}_m(1)}^{\overline{r}_m} = \overline{PK}_1 \\
C_2 \; PK_1 & PK_{\overline{\pi}_1(2)}^{\overline{r}_1} & PK_{\overline{\pi}_2(2)}^{\overline{r}_2} & \cdots & PK_{\overline{\pi}_m(2)}^{\overline{r}_m} = \overline{PK}_2 \\
\vdots & \vdots & \vdots & & \vdots \\
C_n \; PK_n & PK_{\overline{\pi}_1(n)}^{\overline{r}_1} & PK_{\overline{\pi}_2(n)}^{\overline{r}_2} & \cdots & PK_{\overline{\pi}_m(n)}^{\overline{r}} = \overline{PK}_n \\
g & g^{\overline{r}_1} & g^{\overline{r}_2} & \cdots & g^{\overline{r}_m} = h_C \\
\end{array}
$$

Fig. 1: Using exponentiation mixnet to generate pseudonyms. All the terms within the box are published on the bulletin board.

### 4.3 Intermezzo: An attack and a fix on Remark!'s mixnet

Before proceeding further, we have to comment on a finding that we discovered while reflecting on the role of the mixnet in preserving anonymity and unlikability: a new attack against the implementation of Haenni-Spycher's mixnet used in Remark! that compromises anonymity and unlinkability, which we have to fix before proceeding further. We describe an attack that allows the attacker to link any public key to its corresponding pseudonym. Let $L = \{g^{t_1}, \ldots, g^{t_n}\}$ be a list of $n$ values where $g$ is the generator of a multiplicative subgroup $G_q$ of prime order $q$ and $t_i \in Z_q$, $1 \leq i \leq n$.

Let us assume that $E$ is a party that receives $L$ and is requested to send it to the mixnet. We show that malicious $E$ is capable of deanonymizing an arbitrary element of $L$ by adding an additional input to the list. $E$ first chooses a random number $s \in Z_q$ and computes $g^s$. Then, she selects the pseudonym she wants

Secure Internet Exams Despite Coercion        7to deanonymize $g^{t_i} \in L$, inserts $g^{t_i}.g^s$ into $L$ and sends it to the mixnet. The mixnet returns $L' = \{L'_1, \ldots, L'_{n+1}\} = \{g^{t_1 \cdot r}, \ldots, g^{t_n \cdot r}, (g^{t_i}.g^s)^r\}$ as output and publicly publishes $g^r$. Now, $E$ computes $(g^r)^s$ and searches in $L'$ for the element $g^{rs}.L_j$. That element is the pseudonym that links to $t_i$.

The attack exploits the fact that $g^{rs}.g^{t_i \cdot r} = (g^{t_i}.g^s)^r$. It lets $E$ learn $L'_j = g^{t_i \cdot r}$ which is the corresponding output for the chosen input $g^{t_i}$ without $E$ knowing the secret element $t_i$. If $E$ runs it for each element, in $O(n^2)$ time, he can learn all the $n$ links between $L$ and $L'$.

It is worth to stress that if the mixnet has a procedure to check the eligibility of the input public keys run by $E$, the attack is possible if $E$ is untrustworthy, which is usually a health assumption (*i.e.,* better not to have trusted parties): $E$ can violate this check and inputs a manipulated public key. If we assume that the mixnet has no eligibility check feature for inputs, then the attack can even be done by any external attacker who only knows a member from $L$.

The attack mentioned above is independent of the system in which the mixnet is used. Any protocol that relies on the Haenni-Spycher's exponentiation mixnet is vulnerable to this injection attack. In Remark!, exam authority can launch this attack to learn the link between public keys and their associated pseudonyms that violates *Anonymous Marking* and *Anonymous Examiner*.

### 4.4  A Mixnet resilient to Injection Attacks

To secure the mixnet against our linkability attack, one possible solution is to prevent the injection of biased public keys into the mixnet. We propose an *Injection-Resistant Exponentiation Mixnet* setup based on Haenni-Spycher's mixnet, where the input public keys should be accompanied by their ZKPKs for their corresponding private keys. The rest of the structure remains unchanged.

With this simple fix, the mixnet only accepts a public key that is associated with a verified ZKPK and aborts otherwise.

## 5  Security Formal Analysis

Having fixed the mixnet construction, a question remains open: does Remark! with Injection-Resistant Exponentiation Mixnet (IRemix) satisfy *Anonymous Submission* and *Single-Blindness* even if we assume that the parties can coerce one another? If that were true, *Anonymous Submission* and *Single-Blindness* could be simply achieved by ensuring unlinkability. We could even suspect that *Anonymous Submission* and *Single-Blindness* are implied by *Anonymous Marking* and *Anonymous Examiner*. Fortunately, this is not the case, as we argue later in this section, after we set up the formal framework where we perform our security analysis.

*Formalizing e-Exams.* Since we aim at a formal verification, we need to model an e-exam protocol in a formal language. We choose the applied π-calculus (Abadi and Fournet, 2001) for the task, and we refer to the strategy advanced by Dreier *et al.* . in (Dreier et al., 2014), which we remind here.

**Definition 1 (E-exam protocol)** *(Dreier et al., 2014). An* e-exam protocol *is a tuple* $(C, E, Q, A_1, \ldots, A_l, \tilde{n}_p)$, *where $C$ is the process executed by the candidates, $E$ is the process executed by the examiners, $Q$ is the process executed by the question commitee, $A_i$'s are the processes executed by the authorities, and $\tilde{n}_p$ is the set of private channel names.*

Note that all candidates and all examiners execute the same process, but with different variable values, *e.g.,* keys, identities, and answers.

**Definition 2 (E-exam instance)** *(Dreier et al., 2014). An* e-exam instance *is a closed process* $EP = \nu\tilde{n}.(C\sigma_{id_1}\sigma_{a_1}|\ldots\ |C\sigma_{id_j}\sigma_{a_j}|E\sigma_{id'_1}\sigma_{m_1}|\ldots |E\sigma_{id'_k}\sigma_{m_k}|Q\sigma_q|A_1\sigma_{dist}|\ldots|A_l)$, *where $\tilde{n}$ is the set of all restricted names, which includes the set of the protocol's private channels; $C\sigma_{id_i}\sigma_{a_i}$'s are the processes run by the candidates, the substitutions $\sigma_{id_i}$ and $\sigma_{a_i}$ specify the identity and the answers of the $i^{th}$ candidate respectively; $E\sigma_{id'_i}\sigma_{m_i}$'s are the processes run by the examiners, the substitution $\sigma_{id'_i}$ specifies the ith e xaminer's identity, and $\sigma_{m_i}$ specifies for each possible question and answer pair the corresponding mark; $Q$ is the process run by the question committee, the substitution $\sigma_q$ specifies the exam questions; the $A_i$'s are the processes run by the exam authorities, the substitution $\sigma_{dist}$ determines which answers will be submitted to which examiners for grading. Without loss of generality, we assume that $A_1$ is in charge of distributing the copies to the examiners.*

Definition 2 does not specify whether the examiners are machines or humans. For the purpose of our model this distinction is not necessary; it is sufficient that an examiner attributes a mark to a given answer.

Note that $Q$ and $A_1$ could coincide if for instance there is only one authority $A$, in that case we can write simply $A\sigma_q\sigma_{dist}$ instead of $Q\sigma_q|A_1\sigma_{dist}$.

*Model Checking and equational theories.* We use ProVerif (Blanchet, 2001) for the analysis of e-exam protocols. ProVerif allows one to analyze reachability and equivalence-based properties in the symbolic attacker model. We chose ProVerif mainly because it has been extensively used to analyze exam protocols (Giustolisi, 2018), hence we could easily check formerly defined security properties for exams. ProVerif is also one of the few tools that allows for automated analysis of privacy properties using observational equivalence; therefore, we can check *Anonymous Submission* and *Single-Blindness* in our protocol automatically. The input language of ProVerif is the applied $\pi$-calculus (Abadi and Fournet, 2001), which the tool automatically translates to Horn clauses. Cryptographic primitives can be modeled by means of equational theories. An equational theory $E$ describes the equations that hold on terms built from the signature. Terms are related by an equivalence relation $=$ induced by $E$. For instance, the equation $dec(enc(m, pk(k)), k) = m$ models an asymmetric encryption scheme. The term $m$ is the message, the term $k$ is the secret key, the function $pk(k)$ models the public key, the term *enc* models the encryption function, and the term *dec* models the decryption function. The list of all equational theories used to model the Remark! protocol can be found in (Dreier et al., 2014).

### 5.1  Threat Model and Formalization of the New Properties

We let the attacker read all public data on the bulletin board and impersonate misbehaving parties, including an unbounded number of dishonest examiners when checking *Anonymous Submission*, and an unbounded number of dishonest candidates when checking *Single-Blindness*. In addition, we check both properties under a coercion scenario, meaning that the coerced candidate (resp. examiner) reveals their secrets before notification.

Privacy properties can be modeled as observational equivalence properties. To model *Anonymous Submission*, we consider two honest candidates and an unbounded number of dishonest examiners, while to model *Single-Blindness*, we consider two honest examiners and an unbounded number of dishonest candidates. We check both *Anonymous Submission* and *Single-Blindness* considering an honest exam authority. We also check *Anonymous Marking* and *Anonymous Examiner* considering a dishonest exam authority.

First, we check whether *Anonymous Submission* holds in our protocol: specifically, that even if two honest candidates swap their answers in two different runs of the protocol then the attacker cannot distinguish the two resulting systems.

**Definition 3 (*Anonymous Submission*)** *An exam protocol ensures Anonymous Submission if any exam process EP, any two candidates $id_1$ and $id_2$, and any two answers $a_1$ and $a_2$.*

$$EP_{\{id_1, id_2\}}[C\sigma_{id_1}\sigma_{a_1}|C\sigma_{id_2}\sigma_{a_2}]|_{\texttt{marked}} \approx_l EP_{\{id_1, id_2\}}[C\sigma_{id_1}\sigma_{a_2}|C\sigma_{id_2}\sigma_{a_1}]|_{\texttt{marked}}$$

The difference between *Anonymous Submission* and *Anonymous Marking* is that the latter considers two honest candidates who swap their secret keys in two different runs. While both properties aim at hiding the link between candidate's key and answer, the definition of *Anonymous Submission* crucially enables a definition of coercion-resistance. To model this, we additionally let the candidates publish their answers and their secret keys on the public channel, and verify that *Anonymous Submission* still holds:

$$EP_{\{id_1, id_2\}}[C\sigma_{id_1}\sigma_{a_1}|C\sigma_{id_2}\sigma_{a_2}]|_{\texttt{marked}} \approx_l EP_{\{id_1, id_2\}}[C'|C\sigma_{id_2}\sigma_{a_1}]|_{\texttt{marked}}$$

where $C'$ is a process such that $C'^{\setminus \texttt{out}(chc, \cdot)} \approx_l C\sigma_{id_1}\sigma_{a_2}$, i.e. $C'$ is the process that acts like one submitting answer $a_2$, but pretends to cooperate with the attacker by revealing their secrets trough channel *chc*.

Then, we check whether *Single-Blindness* holds in our protocol. We check that if two honest examiners swap their marks in two different runs of the protocol, then the attacker cannot distinguish the two resulting systems.

**Definition 4 (*Single-Blindness*)** *An exam protocol ensures Single-Blindness if for any exam process EP, any two examiners $id_1$ and $id_2$, any two marks $m_1$ and $m_2$*

$$EP_{\{id_1, id_2\}} [E\sigma_{id_1}\sigma_{m_1}|E\sigma_{id_2}\sigma_{m_2}] \approx_l EP_{\{id_1, id_2\}} [E\sigma_{id_1}\sigma_{m_2}|E\sigma_{id_2}\sigma_{m_1}]$$

As for *Anonymous Submission*, we also check that *Single-Blindness* holds under examiner coercion. We let the examiners publish their marks and secret keys on the public channel:

$$EP_{\{id_1,id_2\}}\ [E\sigma_{id_1}\sigma_{m_1}|E\sigma_{id_2}\sigma_{m_2}] \approx_l EP_{\{id_1,id_2\}}\ [E'|E\sigma_{id_2}\sigma_{m_1}]$$

where $E'$ is a process such that $E'^{\setminus \mathsf{out}(chc,\cdot)} \approx_l E\sigma_{id_1}\sigma_{m_2}$.

The difference between *Single-Blindness* and *Anonymous Examiner* is that the latter considers two honest examiner who swap their secret keys in two different runs. While both properties aim at hiding the link between examiner's key and mark, the definition of *Single-Blindness* enables the definition of coercion-resistance.

*Major Security Findings.* ProVerif proves that both properties holds in Remark!. However, it finds attacks when checking the coercion cases in Remark!. *Anonymous Submission* does not hold under candidate coercion because the exam authority publishes on the bulletin board the answer of a candidate along with their pseudo public keys (see step 6 in (Giustolisi et al., 2014)). ProVerif shows an attack trace in which a coercer, who knows the secret key of a coerced candidate, can find out the candidate's answer by retrieving their pseudo public key from the output of the exponentiation mixnet (see step 1 in (Giustolisi et al., 2014)). *Single-Blindness* does not hold under examiner coercion in Remark! because, at notification, the mark and the answer are signed with the examiner's pseudo signing key (see step 8 in (Giustolisi et al., 2014)). ProVerif shows an attack trace in which a coercer, can find out which answers have been marked by a coerced examiner by retrieving their pseudo signing key from the output of the exponentiation mixnet (see step 2 in (Giustolisi et al., 2014)). Table 1 summarizes the findings.

## 6  A Secure Coercion Resistant Exam Protocol

We now present our new protocol, C-Rex, whose goal is to guarantee all the properties outlined above. C-Rex has four main phases *Registration*, *Examination*, *Marking*, and *Notification* and one additional phase which is run before the exam begins. We assume that each test consists of at least two questions. We also assume that there is an append-only bulletin board (BB), which records data concerning examiners, candidates and exam authority. The bulletin board is used to publish public parameters and is also needed for verifiability guarantees. Moreover, we assume that each principal has a pair of Elgamal public/private key which used in an Elgamal cryptosystem (ElGamal, 1985). The five phases of C-Rex are depicted in Figure 2 and are as follows:

*Pre-Assignment*: Let us assume $n$ candidates, who participate in the exam, and $m$ examiners. Before starting the exam, exam authority forms $A = \{1,\ldots,n\}$ and then partitions $A$ into $d$ subsets as $A_P = \{A_{P_1},\ldots,A_{P_d}\}$, and labels them as $P = \{P_1,\ldots,P_d\}$. The exam authority sends $A_P$ and $P$ to each examiner through a secure channel. The examiners sign the partitions and send them back to exam authority. The exam authority distributes the signatures among all examiners so that each examiner can check if they have received the same message as the others.

*Registration*: This phase is as the registration phase in Remark!, with the *injection-resistant exponentiation mixnet* we have explained in Section 4. In this phase, we create pseudo public keys (pseudonyms) as outputs from a list of public keys as inputs while preserving the unlinkability between input and output lists. exam authority sends the list of public keys of eligible candidates and examiners and the list of *corresponding zero-knowledge proofs of knowledge (ZKPKs)* of those keys to the mixnet. Each candidate and examiner, during the authorization and eligibility checks, provides the ZKPK of their keys. When receiving the list of public keys *accompanied by their ZKPKs*, the mixnet checks the proofs and generates the output list in the same fashion as in Figure 1.

*Testing*: In this phase, the exam authority signs and encrypts the questions with the candidate's pseudonym and posts them on the bulletin board. The candidate decrypts the test, checks the signature, and answers the questions. Then sends $[Sign_{SK_C, h_C}(quest, ans, \overline{PK}_C)]_{PK_A}$ where $SK_A$, $PK_A$, $quest$, $ans$ and $\overline{PK}_C$ are the private key of exam authority, the public key of exam authority, the answers, and the candidate pseudonym, respectively. Furthermore, exam authority sends the candidate $[Sign_{SK_A} H(quest, ans, \overline{PK}_C, \alpha)]_{PK_A}$ where $H$ and $\alpha$ are, respectively, a secure hash function and a random value generated by exam authority.

*Marking*: In this phase, we introduce a new technique called *Shuffled Answers* to assign the collected tests to the examiners in a way that guarantees *Anonymous Submission* and *Single-Blindness* also in case of coercion. The idea behind this technique is that when the *Marking* phase starts, each examiner receives a test that each (*question*, *answer*) pair belongs to a random candidate. After collecting the tests from the candidates, exam authority forms a matrix named $T$, which consists of all candidates' (*question*, *answer*) pairs. Then it chooses a secure permutation matrix named $\Pi$ and applies it to $T$ to form a new matrix $T^\pi$, that is $\Pi(T) = T^\pi$. The following transform mathematically expresses the shuffling procedure. Here, if we assume $k$ questions and $n$ candidates, $(q, a)_{(i,j)}$ is the answer of candidate $j$ to question $i$.

$$T = \begin{bmatrix} (q,a)_{1,1} & \cdots & (q,a)_{1,n} \\ \vdots & & \vdots \\ (q,a)_{k,1} & \cdots & (q,a)_{k,n} \end{bmatrix} \xrightarrow{\Pi} \begin{bmatrix} (q,a)_{(1,\pi^1_{(1)})} & \cdots & (q,a)_{(1,\pi^1_{(n)})} \\ \vdots & & \vdots \\ (q,a)_{(k,\pi^k_{(1)})} & \cdots & (q,a)_{(k,\pi^k_{(n)})} \end{bmatrix} = T^\pi$$

Let us define $T$ and $T^\pi$ as sets of $n$ vectors $T = [V_1, \ldots, V_n]$ and $T^\pi = [V_1^\pi, \ldots, V_n^\pi]$, respectively. $V_j$, $1 \leq j \leq n$ shows the test of candidate $j$, while $V_j^\pi$, $1 \leq j \leq n$ represents a new test whose each question belongs to a random author. exam authority signs $T^\pi$ with its private key and publishes it on the bulletin board in front of its public key. Then exam authority randomly assigns each examiner a member of $P$ and posts this assignment on the bulletin board. Therefore, each examiner sees a label in front of their pseudonym on the bulletin board that means which subset they should mark. Let us name $P_E$ the label assigned to the examiner $E$. $E$ grades the corresponding assigned vectors and, for each, sends the message $[Sign_{SK_E, h_E}(M_j^\pi, A_{P_E})]_{PK_A}$ to exam authority where $M_j^\pi$ is the vector of marks associated with $V_j^\pi$ and $j \in A_{P_E}$.

*Notification*: Let $S_j^\pi = Sign_{SK_E, h_E}(M_j^\pi, A_{P_E})$. When exam authority receives all messages from all examiners, first it checks if examiners marked the correct assigned subsets, and then it constructs two $S^\pi = [S_1^\pi, \ldots, S_n^\pi]$ and $M^\pi = [M_1^\pi, \ldots, M_n^\pi]$ matrices. Then, it applies the inverse permutation $\Pi^{-1}$ to these matrices and generates $S = [S_1, \ldots, S_n]$ and $M = [M_1, \ldots, M_n]$, respectively. $M$ is the matrix of final marks and each column of it, which means $M_j$, is a vector showing the marks for the $j^{th}$ candidate's test. Furthermore, $S$ represents a notification matrix that actually includes the marks signed by the eligible examiners. exam authority signs and encrypts each column of $S$ with the pseudonym of the corresponding candidate and then posts the output on the bulletin board. Finally, the mixnet servers reveal their secret exponents that are used to anonymize the candidates. Therefore, the anonymity of the candidate is revoked and exam authority can register the marks. Furthermore, exam authority reveals the secret element $\alpha$ at the end of the exam.

Let us now assume that the coercer has asked a candidate to reveal her private key and her submitted test. The coerced candidate first looks at $T^\pi$ and, from each row, picks an arbitrary pair of *(question,answer)*. Then, she reveals her real private key, as well as the set of *(question,answer)* pairs chosen. We claim that the coercer cannot distinguish whether the candidate has demonstrated her real test or a fake one. Instead, let us assume an examiner $E_j$, $1 \leq j \leq d$ who is coerced by a candidate and is supposed to mark the tests labeled $P_r \in P$. In addition to $P_r$, $E_j$ marks $P_{r'}$ where $P_{r'} \in P$ and $r \neq r'$. Now, if the coercer asks $E_j$ to reveal her secret keys, $E_j$ pretends that she has marked only the tests labeled $P_{r'}$. Our assertion is that the coercer cannot distinguish whether $E_j$ lies about her assigned partition.

*Security Formal Analysis* ProVerif proves that our protocol meets both *Anonymous Submission* and *Single-Blindness*, including the coercion alternatives. Our protocol resists the attack on candidate coercion seen in Remark! because the exam authority re-randomizes the pseudo public key of the candidate. Single-Blindness is proved also under examiner coercion because, differently from Remark!, a coerced examiner can lie to a coercer by claiming that the examiner is marking a different partition of tests.

In addition to *Anonymous Submission* and *Single-Blindness*, we prove in ProVerif that our protocol meets all the original properties meet by Remark!, namely, anonymous marking, anonymous examiner, test answer authentication and examiner authentication. Table 1 summarizes the findings.

Table 1 highlights the dishonest parties in our analysis. For *Mark Privacy*, $C^*$ and $E^*$ are, respectively, all candidates and examiners except the candidate and examiner concerned. Collusion is possible when at least two parties misbehave.

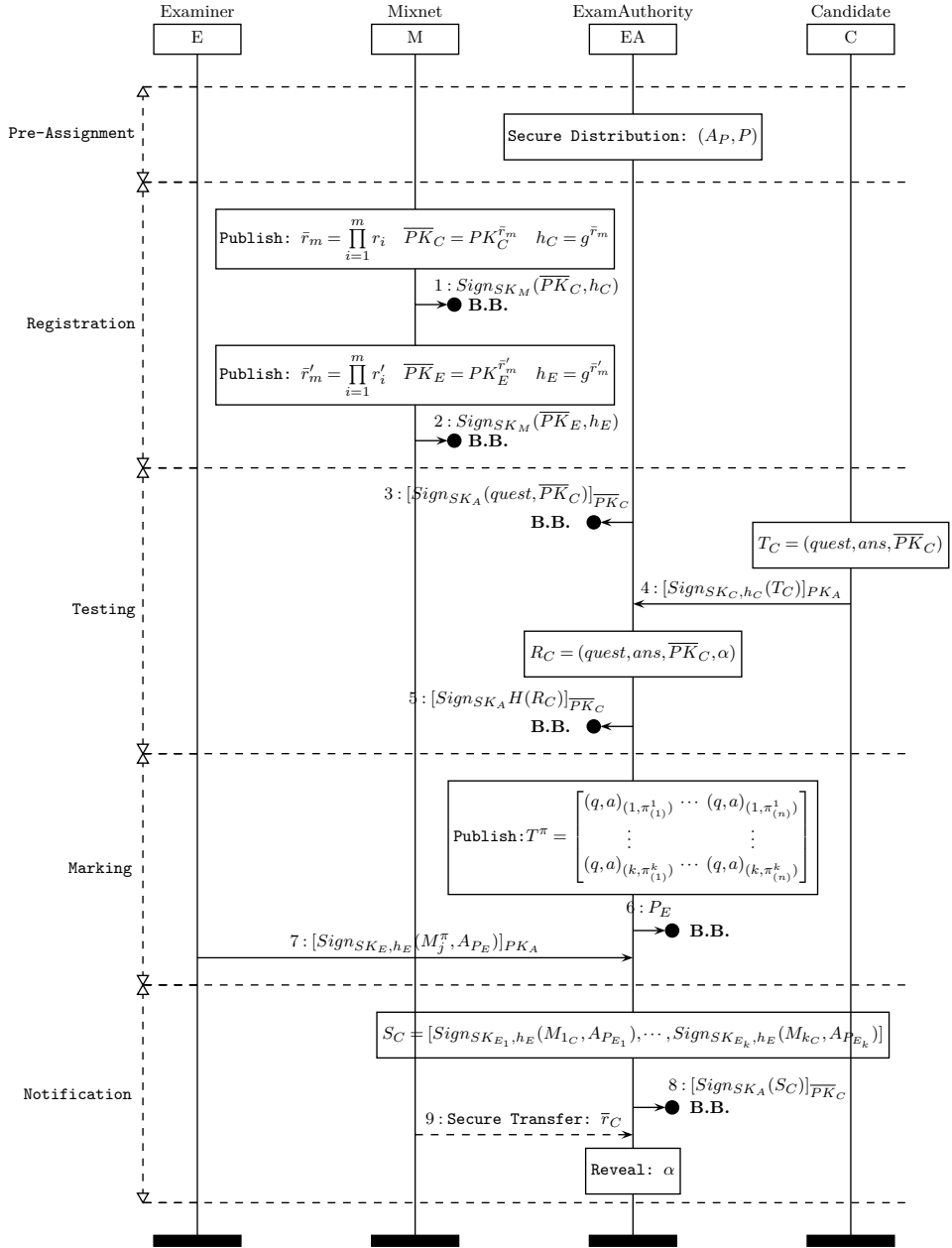

Fig. 2: The message sequence chart of C-Rex

14      Mohammadamin Rakeei , Rosario Giustolisi, and Gabriele Lenzini

| Properties | **Remark!** (with IRemix) | **C-Rex** | Threat Model |
|---|:---:|:---:|---|
| Answer Authentication | ✓ | ✓ | Dishonest E |
| Examiner Authentication | ✓ | ✓ | Dishonest C |
| Mark Privacy | ✓ | ✓ | Dishonest ($C^*$,$E^*$) |
| Anonymous Marking | ✓ | ✓ | Collusion(EA,E) |
| Anonymous Examiner | ✓ | ✓ | Collusion(EA,C) |
| **Anonymous Submission** | ✓ | ✓ | Dishonest E |
| **Single-Blindness** | ✓ | ✓ | Dishonest C |
| **Anonymous Submission** | ✗ | ✓ | Coercion(E) |
| **Single-Blindness** | ✗ | ✓ | Coercion(C) |

Table 1: Synthesis of the findings of the formal analysis under different threats.

Therefore, *Anonymous Marking* and *Anonymous Examiner* are discussed in the collusion threat model with respectively *EA* & *E*, and *EA* & *C* being dishonest. *Anonymous Submission* and *Single-Blindness* are studied under both coercion and non-coercion scenarios.

## 7   Conclusion

The use of online exams, which peaked during the pandemic, raises issues of security and privacy, since it is easier to cheat when exams are held in remote.

In certain sectors, like in e-voting, security and privacy are already well-established subjects of research. Various academic and industrial collaboration activities have been established, which now support the sectors with ideas, prototypes, forums for discussions, and projects. Compared to e-voting, electronic exams seem underrepresented. One could discuss whether what is at stake in e-voting, aka one important cog in the democracy process, is more important than the quality of assessment of skills and knowledge of people, but recent events show that the government attention to a fair and honest assessment is not second to anything. At the time of writing, giant companies like Ernst & Young, admitted their employees had cheated on ethics exams, an act that cost them a fine of a hundred million USD [10]. This episode is not isolated. In March 2022, PwC has been sanctioned about one million USD for "having faulty quality control standards that allowed more than 1,200 professionals to cheat on internal training courses" [11]. In a different domain, many universities are struggling to achieve robust online assessment systems, where at stake is the trust that we have in the general reward strategy of our educational systems. It is clear that to adhere to a code of conduct is not a sufficient guarantee for an honest outcome and that we need better, more secure, and private by-design exam protocols.

---

[10] *id. at* 4

[11] Soyoung Ho, "Canada for Widespread Employee Cheating on Internal Tests", Thomson Reuters, 2 March 2022, last access 2022/07

The presented work studies an important class of security requirements missing in the state-of-the-art about exams: coercion resistance. Following this main goal, we obtained several important achievements. First, we formally defined two new properties, *Anonymous Submission* and *Single-Blindness*, which allow one to reason about the phenomenon of coercion in online exams. Thanks to this effort, we find that a state-of-the-art protocol is not coercion resistant in the sense we describe with our properties. Coercion resistance requires a different cryptographic approach, and we propose a new cryptographic protocol Coercion-Resistant Electronic Exam (C-Rex). To our knowledge, it is the first coercion-resistant e-exam protocol: unlike Remark!, it guarantees *Anonymous Submission* and *Single-Blindness* even if the parties can coerce one anotherinto revealing secrets. C-Rex implements a new secure exponentiation mixnet (we call it Injection-Resistant Exponentiation Mixnet (IRemix)) which is also an original although orthogonal contribution of this work. In fact, while investigating the security of a state-of-the-art e-exam protocol, Remark!, we found a novel linkability attack on the Haenni-Spycher's mixnet, the main building block of Remark!, which completely broke the claimed unlinkability property of this mixnet. As a countermeasure against this attack, we proposed IRemix as an injection-resistant exponentiation mixnet based on the structure of Haenni-Spycher. It is used in our C-Rex scheme.

Although we proposed IRemix as the secure alternative for the Haenni-Spycher's structure, its security is just informally discussed and it is an important further work to formally verify the security of IRemix. Furthermore, in the *Pre-Assignment* phase, we proposed a simple protocol which assures that all parties have received the same set of partitions. It is desirable for future research to formally verify this protocol.

Future studies could also investigate the security of the systems that used Haenni-Spycher's mixnet as an identity mixer, against the linkability attack that we found. The first one could be the Haenni-Spycher's e-voting protocol (Haenni and Spycher, 2011) which seems vulnerable to this attack if the election authority is dishonest. Other schemes that might be prone to the attack are (Dubuis et al., 2013), (Locher and Haenni, 2014), (Ryan, 2016), and (Haenni and Koenig, 2013). Since our formal verification of the C-Rex protocol in this work was just about the authentication and privacy properties, another direction for future research could be investigating verifiability guarantees of the C-Rex e-exam.